\documentclass[aps,prl,twocolumn,superscriptaddress, amssymb]{revtex4}

\usepackage{graphicx}
\usepackage[ocgcolorlinks]{hyperref}

\begin{document}

\title{Laser micro-fabrication of concave, low-roughness features in silica}

\author{D. Hunger}
\affiliation{Max-Planck-Institut f{\"u}r Quantenoptik and Fakult{\"a}t f{\"u}r Physik der Ludwig-Maximilians Universit{\"a}t,
  Schellingstra{\ss}e~4, D~80799~M{\"u}nchen, Germany}
\author{C. Deutsch}
\affiliation{Laboratoire Kastler Brossel, ENS/UPMC-Paris 6/CNRS, 24
  rue Lhomond, F-75005 Paris, France}
\author{R. J. Barbour}
\affiliation{School of Engineering and Physical Sciences, Heriot-Watt University, Edinburgh EH14 4AS, UK}
\author{R. J. Warburton}
\affiliation{Department of Physics, University of Basel,
Klingelbergstrasse 82, CH-4056 Basel, Switzerland}
\author{J. Reichel}
\email{jakob.reichel@ens.fr}
\affiliation{Laboratoire Kastler Brossel, ENS/UPMC-Paris 6/CNRS, 24
  rue Lhomond, F-75005 Paris, France}

\date{\today}

\begin{abstract}
  We describe a micro-fabrication method to create concave features with ultra-low
  surface roughness in silica, either on the end facets of optical fibers or on flat substrates. The machining uses a single focused CO$_2$ laser
  pulse. Parameters are chosen such that material is removed by
  thermal evaporation while simultaneously producing excellent surface
  quality by surface tension-induced movement in a low-viscosity melt
  layer.  A surface roughness $\sigma\sim0.2\,$nm is regularly obtained. The
  concave depressions are near-spherical close to the center with
  radii of curvature between $20$ and $2000\,\mu$m.  The method allows the
  fabrication of low-scatter micro-optical devices such as mirror
  substrates for high-finesse cavities or negative lenses on the tip
  of optical fibers, extending the range of micro-optical components.
\end{abstract}


\maketitle

CO$_2$ laser light is a powerful tool for the surface treatment of fused
silica. Strong absorption of the $10.6\,\mu$m light within the first
few $\mu m$ enables controlled surface melting using moderate
intensities. 
A central benefit of such treatment is that
surface tension in the molten layer smoothens out surface roughness up to
a scale comparable to the thickness of the melt layer. This was
recognized as an efficient way to polish optical surfaces
\cite{Laguarta94,Nowak06}. Owing to the surface-minimizing effect
of the surface tension, melting of large-scale domains which penetrate
considerably into the volume will result in the formation of convex
structures. While this behaviour has been successfully employed for the
fabrication of convex micro-optical structures
\cite{Paek75,Wakaki98,Vernooy98,Armani03}, 
it is undesired in the context of surface polishing.
At higher intensities, material removal
by evaporation becomes important. In the context of optical surfaces,
evaporation has mainly been studied as an unwanted side-effect of damage
repair \cite{Feit02,Mendez06} and polishing
\cite{Nowak06}. Nevertheless, it holds promise for
micro-machining of optical components \cite{Markillie02}.

Here we describe micromachining of concave profiles on silica surfaces
in a simple process using a single laser pulse. We work in a regime
where surface evaporation is dominant and shapes a depression, while
melting is restricted to a thin layer that smoothens the surface on a
short scale. This avoids the need for scanning techniques, special
atmospheres, preheating, or polishing pulses. The method allows
fabrication of microscopic concave surfaces on the tip of optical
fibers as well as on bulk fused silica substrates. It was originally
developed to realize high-finesse fiber Fabry-Perot cavities (FFPs)
that enabled the first cavity quantum electrodynamics (CQED)
experiments with Bose-Einstein condensates \cite{Colombe07}, and very
recently has been used with single organic molecules
\cite{Toninelli10} and quantum dots \cite{Barbour11}. More details on
these FFPs can be found in \cite{Hunger10b, Muller10}.  Here, we
present an investigation of the physical process leading to the
formation of these ultrasmooth concave depressions.

\begin{figure}[tb]
\includegraphics[width=\columnwidth]{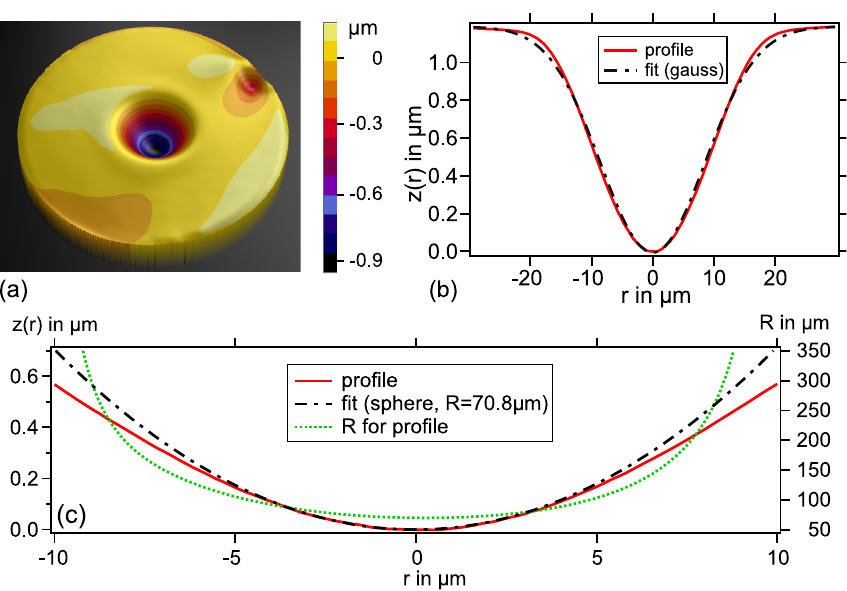}%
\caption{\label{Img:Profile} (a) Surface profile of a machined fiber
  measured by the optical profilometer. Beam
  parameters are $P=420\,$mW, $\tau=27\,$ms and $w=28\,\mu$m. (b) Cut
  through the center of the profile shown in (a) (red solid line) and
  its fit to a Gaussian (black dash-dotted line). (c) Central part of
  the same data (red solid line) along with a circle
  fitted to the center
  yielding $R=70.8\,\mu$m (black dash-dotted line). Also shown is
  the local radius of curvature as calculated from a
  high-order polynomial fit to the data (green dotted line).}
\end{figure}

We use an RF-pumped CO$_2$ laser (Synrad Firestar v20) with a
nominal power of 20\,W. Pumping is pulsed at 20\,kHz, the average
power $P$ being controlled via the duty cycle. Laser power
fluctuations are minimized by reducing unwanted back-reflections to
the laser with a quater-wave plate and polarizer, and we obtain energy
fluctuations of 1.2\% for a 4\,ms pulse train. The beam is focussed
onto the sample surface. To position the sample with respect to the focal
waist, we use a long working distance microscope that allows us to inspect the
surface via a dichroic mirror. We estimate the alignment precision to
2\,$\mu$m transversally and 7\,$\mu$m along the beam axis, limited by
the depth of field of the microscope.

Our substrates are metal-coated, single-mode or gradient-index
optical fibers (Oxford Electronics) with 125 and 200\,$\mu$m
cladding diameter, and 2\,mm thick fused silica glass plates
(Hellma QS). The fibers were not cleaned after cleaving; the plates
were roughly cleaned with ethanol.  After machining, surfaces were
analyzed with several methods. Surface roughness $\sigma_\mathrm{rms}$
is evaluated by atomic force microscopy (Dimension 3100 AFM)
and optical loss measurements, discussed further below. The overall
shape is measured by optical interferometric profiling
(Fogale Micromap 3D).  Figure \ref{Img:Profile} shows a
typical machined surface. We observe profiles with an approximately
Gaussian shape. The geometrical properties used for characterization
are the central radius of cuvature $R$, the total depth $t$ and the
structure diameter $d$. We evaluate $R$ from a polynomial fit to the
center of the structure. We define $t$ as the height difference
between the maximum and the minimum elevation, and
$d$ as the distance between the turning points
from convex to concave curvature of a high-order polynomial fit.

\begin{figure}[tb]
\includegraphics[width=\columnwidth]{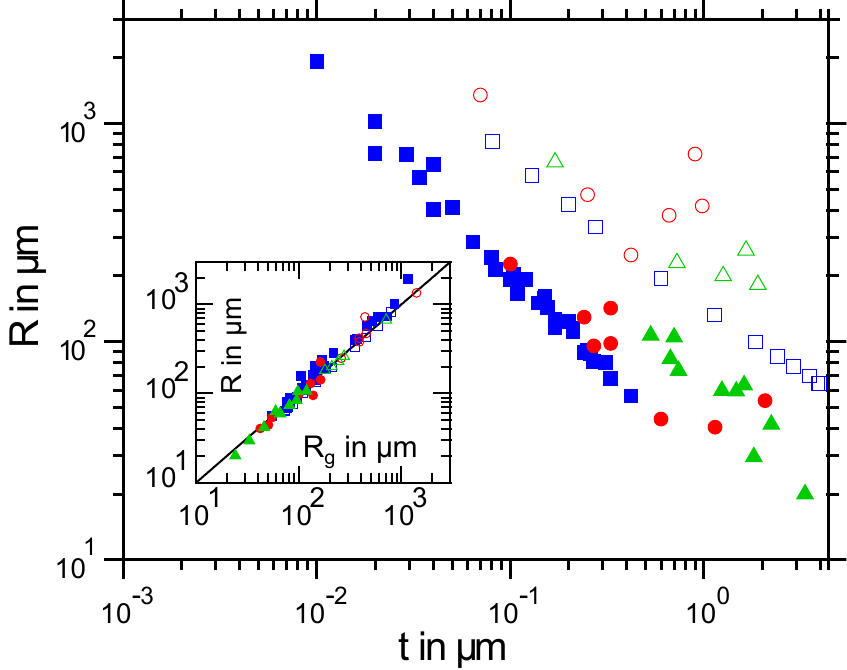}%
\caption{\label{Img:Geometry} Experimentally realized geometries: $R$
  as a function of $t$. Beam powers and waists:
  $\blacksquare~600\,\mathrm{mW},\ 26\,\mu$m. 
  $\square~1.85\,W,\ 73\,\mu$m. 
  {\large{$\bullet$}} 373\,mW, 27\,$\mu$m. 
  {\large{$\circ$}}  0.9\,W, 63\,$\mu$m. 
  $\blacktriangle$~540\,mW, 28\,$\mu$m. 
  $\triangle$~575\,mW, 63\,$\mu$m. 
  Data for a given parameter set were taken by varying $\tau$; $\tau$
  increases from left to right.
  Squares: bulk material, circles: $\varnothing 125\,\mu$m fiber,
  triangles: $\varnothing 200\,\mu$m fiber. Inset: The measured $R$
  agrees well with the value $R_g$ calculated from $d$ and
  $t$. (Straight line: $R=R_g$.)}
\end{figure}

We machined a large set (several hundreds) of fibers, and a series of
glass plates, with a broad range of parameters.  Laser powers were
between $P=300\,$mW and 2\,W at the sample surface and pulse trains had
durations $\tau= 4$ to 120\,ms (consisting of 80 to 2400 pulses). Beam
waists were between $w=21$ and $93\,\mu$m.  With these parameters,
the resulting structures are $t = 0.01\ldots 4\,\mu$m deep and have
diameters of $d = 10\ldots 60\,\mu$m, leading to radii of curvature
$R= 20\ldots 2000\,\mu$m.  Fig.~\ref{Img:Geometry} shows how the results
depend on the laser parameters. For increasing $\tau$, the structures become
deeper and $R$ becomes smaller. For increasing $w$, both $d$ and $R$
increase also.
The measured $R$ agrees well with the value $R_g=d^2/(8t)$ expected
for a gaussian shape (inset in Fig.~\ref{Img:Geometry}): of the three
parameters $\{d,t,R\}$, only two are independent and fully determine
the third.

The fiber data show a spread for fixed parameters of up to $13\%$ in
the profile depth, $23\%$ in the profile diameter and $13\%$ in the
radius of curvature. Probable causes for this spread are the
fluctuations in the laser energy and the limited positioning precision
along the optical axis.

\begin{figure}[tb]
\includegraphics[width=\columnwidth]{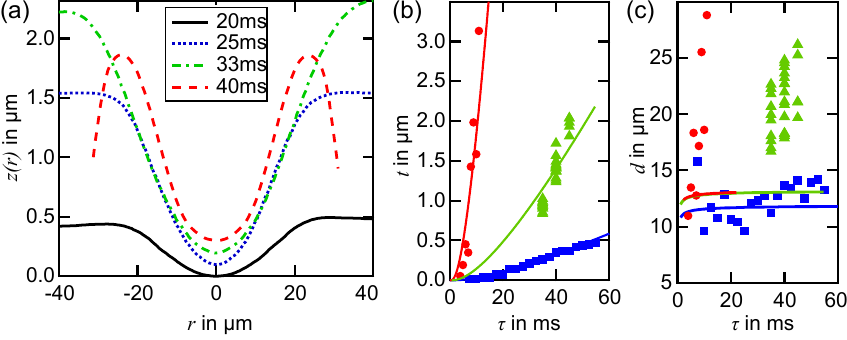}
\caption{\label{Img:depth_upwelling} (a) Profile as a
  function of $\tau$. Shown are cuts through the center of $\varnothing
  125\,\mu$m fibers for $P=572\,$mW, $w=51\,\mu$m. The curves are
  vertically offset by 100\,nm for readability.
  (b) Central depth and (c) diameter of the depression as a function
  of $\tau$ for similar beam parameters on different targets:
  \large{$\bullet$}\small~$\varnothing125\,\mu$m
  fibers. $\blacktriangle~\varnothing200\,\mu$m
  fibers. $\blacksquare~2\,$mm thick fused silica plate. Solid lines:
  Fit to the described model, where $\kappa$ is used as fitting
  parameter to the data in (b). We find $\kappa = 2.05\,$W/mK for the
  $125\,\mu$m fibers, $\kappa=2.0\,$W/mK for the $200\,\mu$m fiber,
  and $\kappa = 2.55\,$W/mK for the glass plate. The same parameters
  are used for the model curves in (c), yielding a reasonable
  agreement for the plate, but not for the fibers. Beam parameters:
  \large$\bullet$ \small 600\,mW,
  26\,$\mu$m. $\blacktriangle$~540\,mW,
  28\,$\mu$m. $\blacksquare$~600\,mW, 26\,$\mu$m.  }
\end{figure}

The possible range of laser parameters for the fabrication of concave
profiles is restricted by two physical limits. For very short pulses
at high intensity, melting and resolidification occur faster than the
surface smoothing process \cite{Nowak06}. (Our parameters do
not reach this limit.) For very long
pulses, the melt layer extends far into the volume and leads to global
contraction into a convex shape. This regime is announced by the
formation of a bulge around the depression at our longest pulse
durations (red dashed curve in Fig.~\ref{Img:depth_upwelling} (a)).
The bulged depressions deviate from a Gaussian shape, becoming
shallower and broader.

Some insight into the evaporation process can be gained from a simple
model in which the surface
temperature profile is used to calculate the ablation rate.  The local temperature is calculated considering the
absorption of a laser beam (intensity profile
$I(r)=I_0\exp(-2r^2/w^2)$) by the surface of a half-space filled with
a homogeneous medium. The medium is characterized by its absorption
coefficient $A$, thermal conductivity $\kappa$, and thermal
diffusivity $D$. (Numerical values for fused silica at room temperature
are $A=0.85$, $\kappa=1.38\,\textrm{W}/\textrm{mK}$, and $D=7.5 \times
10^{-9}\,\textrm{m}^2/\textrm{s}$.) The resulting surface temperature
profile is obtained from the two-dimensional heat flow equation
\cite{Allmen95,Feit02}:
\begin{equation}\label{eq:TempDist}
  T(r,\tau)=\frac{A w^2 I_0}{2 \sqrt{\pi}}\frac{\sqrt{D}}{\kappa}\int^{\tau}_{0}\frac{e\,^{\frac{-r^2}{w^2/2+4D\tau'}}}{\sqrt{\tau'}(w^2/2+4D\tau')}\,d\tau'.
\end{equation}
The temperature in the center can be evaluated analytically,
$T(0,\tau)={AI_0 w}/(\sqrt{2\pi}\kappa)\arctan\sqrt{{\tau}/({w^2/8D})}$.
The material removal rate (velocity of the evaporation
front) is related to the local temperature as
$v(r,\tau)=v_0e^{-U/k_BT(r,\tau)}$, where $U=3.6\,$eV is the latent heat of
evaporation per atom and $v_0=3.8 \times 10^5\,$cm/s \cite{Feit02}.
It becomes significant above $\sim 2000^{\circ}$C and has a very
strong temperature dependence in this temperature range.
This simple expression neglects all non-linearities as well as the
boundary condition set by the finite size of the fiber that modifies
the transverse heat conduction.

For short pulses (no significant bulge), the depth of the structures
is reproduced well by this model (Fig.~\ref{Img:depth_upwelling}(b)).
We treat $\kappa$ as a free fitting parameter on account of its strong
temperature dependence; the resulting $\kappa$ values are in the range
of published values \cite{Yang09} for hot silica. (The small
difference between the $\kappa$ values for the two fibers is
not significant considering the uncertainties of the power and waist
measurements.) The observed profile diameters for plate machining are
in reasonable agreement with the model.  For fiber machining, however,
the profile diameters are systematically larger than predicted by the
model and also show a different parameter dependence
(Fig.~\ref{Img:depth_upwelling}(c)). A possible explanation is the
boundary condition set by the cylindrical geometry of the fiber. The
restriction of the transverse heat flow causes heat accumulation and
thus leads to an enlarged melt and evaporation area.

\begin{figure}[t]
\includegraphics[width=\columnwidth]{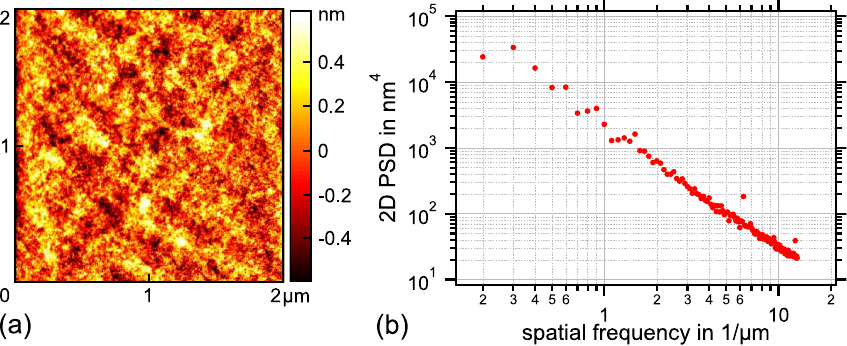}%
\caption{\label{Img:PSD} (a) AFM measurement of a $2\times2\,\mu$m$^2$ area in the center of a laser machined area. Shown is the surface elevation after substracting a fitted polynomial. (b) The 2D PSD of a $5\times5\,\mu$m$^2$ AFM scan for the remaining height elevation.}
\end{figure}

The surface roughness is determined by AFM measurements at different
positions with scan areas of $0.5, 2$ and $5\,\mu$m size to avoid
artefacts and to minimize noise. To distinguish between micro
roughness, which is important for the optical quality, and the larger
scale shape which determines the imaging properties, the deviation
from a polynomial fit from each scan line is used for roughness
calculations. All measurements show comparable roughness values of
$\sigma =0.24\,$nm rms. Fig.~\ref{Img:PSD} shows the residual from
this fit and a two dimensional power spectral density calculated from
the data (2D PSD).  Reference measurements on mica sheets enable us to
estimate the noise of the measurement to be
$\sigma_{\textrm{\small{noise}}}=0.1\,$nm. Correcting for this yields
$\sigma = 0.22\,$nm rms for the laser-machined structures, a
value approaching that of superpolished optics. Scatter loss can be
estimated \cite{Bennett92} by $S\approx (4\pi \sigma
/\lambda)^2$, which predicts $S\approx 11\,$ppm at $\lambda=830\,$nm
for this roughness. For two-fiber cavities with state-of-the-art
coatings, we have recently obtained a cavity finesse of 
$\mathcal{F}= 100,000$, compatible with the surface roughness
result within the uncertainty of the measured cavity transmission.

While traditional mirror superpolishing methods are restricted to
large radii of curvature in the centimeter to meter range
\cite{Hood01}, the method described here achieves
extremely small radii of curvature, and allows a wide range of
geometries not limited to the range of data shown here. It thus opens
a previously inaccessible geometry regime for high-quality
micro-optical lenses and mirrors, starting from low-cost standard
substrates. First applications in neutral-atom, quantum dot and
molecule CQED \cite{Colombe07,Volz11,Barbour11,Toninelli10} have already shown
the power of this method, in realizing microscopic, high-finesse optical cavities with small mode
volume. Applications in  cavity optomechanics
and quantum information experiments with ion traps are under way.

\begin{acknowledgments}
We thank J.~Hare and his team (LKB, Paris) for access to their CO$_2$
laser in early stages of this project, D.~Chatenay and his team (LPS,
Paris) for access to their optical profiler and the CENS (Munich) and
INSP (Paris) institutes for access to their AFMs. R.~W. acknowledges
fruitful discussions with Kris M.~Nowak, Howard J.~Baker and Denis
R.~Hall at Heriot-Watt University. This work was supported in part by
the AQUTE Integrated Project of the EU (grant agreement 247687) and by
the Institut Francilien pour la Recherche sur les Atomes Froids (IFRAF).
\end{acknowledgments}


\end{document}